\documentclass[pdflatex,sn-mathphys-num]{sn-jnl}

\usepackage{graphicx,natbib}%
\usepackage{multirow}%
\usepackage{amsmath,amssymb,amsfonts}%
\usepackage{amsthm}%
\usepackage{mathrsfs}%
\usepackage[title]{appendix}%
\usepackage{xcolor}%
\usepackage{textcomp}%
\usepackage{manyfoot}%
\usepackage{booktabs}%
\usepackage{algorithm}%
\usepackage{algorithmicx}%
\usepackage{algpseudocode}%
\usepackage{listings}%
\newcommand{\mat}[1]{\mbox{\boldmath{$#1$}}}

\theoremstyle{thmstyleone}%
\theoremstyle{thmstyletwo}%

\theoremstyle{thmstylethree}%

\begin{document}

\title[Cox Model Predicting Covariate Subject to Right Censoring]{Cox Model Predicting Covariate Subject to Right Censoring}


\author[1]{\fnm{Chen-Yen} \sur{Lin}}
\equalcont{These authors contributed equally to this work.}

\author*[2]{\fnm{Susan} \sur{Halabi}}\email{susan.halabi@duke.edu}
\equalcont{These authors contributed equally to this work.}

\author[3]{\fnm{Taehwa} \sur{Choi}}

\affil[1]{\orgdiv{Oncology Biometrics}, \orgname{AstraZeneca}, \orgaddress{\country{Canada}}}

\affil*[2]{\orgdiv{Department of Biostatistics and Bioinformatics}, \orgname{Duke University}, \orgaddress{\street{2424 Erwin Rd}, \city{Durham}, \postcode{27705}, \state{NC}, \country{USA}}}

\affil[3]{\orgdiv{School of Mathematics, Statistics and Data Science}, \orgname{Sungshin Women's University}, \orgaddress{\state{Seoul}, \country{South Korea}}}


\abstract{Time-to-event endpoints are frequently used as outcomes in oncology and other disease areas where the outcome of interest may not be observed within a predetermined period. Although many analytical methods address the challenges of censoring in outcomes, limited research has focused on censored covariates. Conventional methods such as the complete case (CC) analysis, where data from patients with censored covariates are discarded, suffer from efficiency loss and potential bias due to reduced sample size. Alternatively, imputing censored covariates with a constant value can underestimate variability. Recognizing these limitations, novel estimation procedures within the generalized linear model framework have been proposed, with some research emerging in time-to-event outcomes. In this paper, we investigate the association between progression-free survival and overall survival using a semi-parametric Cox model framework. We modify the Cox model's partial likelihood function to account for censored covariates by replacing the relative risk associated with censored covariates with a weighted average of patients with observed covariates. The performance of the proposed method is demonstrated through simulations and applications to two oncology clinical trials. Results indicate that the proposed method offers improved estimation efficiency and better utilization of available data compared to other  approaches.}

\keywords{Cox model, censored covariate, prognostic model, innovation theorem}



\maketitle

\section{Introduction}
\label{sec1}
Time-to-event endpoints are commonly used in oncology and other disease areas where the outcome of interest may not be fully observed within a pre-specified study period. Although many methods have been developed to address censoring in outcomes, relatively limited research has been dedicated to address the challenges associated with censored intermediate outcomes. Such censoring can occur due to loss to follow-up, patient withdrawal,  study termination, or detection limits,  all of which complicate the accurate estimation and interpretation of intermediate effects \citep{SurvBook}.

There is considerable interest in exploring whether intermediate outcomes, such as the duration until a patient responds to initial treatment can predict overall survival (OS) or whether the duration of pain a patient experiences predicts OS \citep{Halabi08,Halabi09}. For example, studies in metastatic castration-resistant prostate cancer suggest that patients who do not progress shortly after initiating a new treatment tend to experience longer survival compared to those who progress early. Moreover, prior data have shown that time to prostate specific antigen (PSA)$\leq$0.2 is prognostic for OS \citep{Matsubara20,Chowdhury23}.  Similarly, in breast cancer, achieving a complete pathological response correlates with longer OS \citep{Cortazar14}. However, these studies typically dichotomize the intermediate response and evaluate is association marginally, without adjusting for other potential prognostic factors or fully leveraging the full information of the time to event intermediate outcome. For ease of discussion, in this manuscript we refer to the intermediate time to event outcome as a censored covariate.

Analyses of fully observed baseline covariates have traditionally applied standard methods directly, such as the Cox proportional hazards model. In contrast, when an intermediate outcome serves as a covariate but is subject to censoring, the analysis framework becomes more complex. Methods commonly used for baseline covariates, such as complete case (CC) analysis, are not applicable to censored covariates. While CC can yield unbiased estimates under a missing completely at random assumption \citep{MissingBook}, it reduces efficiency by shrinking the effective sample size and may violate the intent-to-treat principle, particularly when the censored covariate is associated with the primary endpoint.  Another historically common approach replaces covariate values with constants, such as the mean or median. Although this retains all participants, it underestimates variability and can produce unstable parameter estimates. These limitations highlight the need for more rigorous methods when intermediate outcomes are modeled as censored covariates in survival analysis.


Recognizing these limitations, some estimation procedures have been proposed within the generalized linear model framework \citep{Tsimikas12,Atem17,Qian18,Matsouaka20} and for time-to-event outcomes \citep{Lee03,DAngelo08,Bernhardt14,Atem19}. For example, \cite{Bernhardt14} developed methods under the accelerated failure time (AFT) model, while \cite{Atem19} proposed a Cox model framework that replaces censored covariate values with their conditional expectations. Although informative, these approaches may fail to fully account for censoring, do not adjust adequately for other prognostic factors, or perform unstably in small samples or trials with heavily censored intermediate outcomes.

The motivation for this work stems from the pressing need to identify reliable intermediate endpoints that can predict overall survival (OS) in oncology clinical trials. OS remains the gold standard endpoint for evaluating treatment benefit, but it often requires lengthy follow-up and large patient cohorts, delaying regulatory decisions and patient access to effective therapies. Intermediate time-to-event endpoints—such as time to progression (TTP) and progression-free survival can be observed earlier, providing the opportunity to assess treatment efficacy sooner and potentially accelerate drug development. 

Although the association between PFS or TTP and OS can be quantified using copula-based correlation \citep{Xie2020}, correlation alone captures only the strength of the association and does not establish a predictive value. A strong correlation does not clarify whether knowledge of the intermediate endpoint meaningfully improves OS prediction for individual patients. Moreover, copula-based methods are typically marginal, failing to adjust for other covariates or account for censoring in the intermediate outcome. Reliance solely on correlation can thus obscure clinically meaningful relationships and misestimate the utility of intermediate endpoints for survival prediction.

To address this gap, we propose modeling the relationship between the intermediate time to event endpoints (such as PFS or TTP) and OS within a semi-parametric Cox framework, which allows us to formally evaluate the predictive contribution of an intermediate time-to-event outcome while adjusting for relevant covariates and properly handling censoring.  Unlike prior approaches that focus on baseline covariates or marginal associations, our method is designed specifically for post-baseline, intermediate time-to-event covariates. It directly evaluates the predictive value of such covariates for OS, adjusting for other prognostic factors and properly accounting for censoring.  

The remainder of the article is structured as follows.  In section \ref{MethodSec} we introduce the notation and describe our modification of the Cox proportional hazards model and its associated partial likelihood function. We evaluate the performance of the proposed estimation procedure via  simulations in Section \ref{SimSec} and apply the method to two clinical trials in oncology in Section  \ref{realdata}. We offer concluding remarks in Section \ref{conclusion}.

\section{Methods}\label{MethodSec}
\subsection{Notation and preliminary modeling}
Suppose $n$ patients are independently observed. For the $i$-th patient, let the outcome of interest be a time-to-event endpoint where $T_i$ and $T^c_i$ are the true survival and censoring time, respectively. Consider a $p+1$-dimensional covariate vector $(W_i~~\mathbf{X}^T_i)^T$, where $W_i$ is the covariate that is subject to right censoring per censoring mechanism $W^c_i$ and $\mathbf{X}_i$ is a $p$-dimensional fully observable covariates. We denote the observed outcome and the covariate as $Y_i=\min(T_i,T^c_i)$ and $Z_i=\min(W_i,W^c_i)$, respectively, with their corresponding censoring indicator $\delta_i=I(T_i\leq T^c_i)$ and $\eta_i=I(W_i\leq W^c_i)$.

In the Cox model framework, the hazard function of $T$ has the form
\begin{equation}\label{coxconcept}
h(T)= h_0(T)\exp[\gamma W+\mat{\beta}^T\mathbf{X}],
\end{equation}
where $h_0(\cdot)$ is an unspecified baseline hazard function, the multiplicative factor $r(\gamma,\mat{\beta})=\exp[\gamma W+\mat{\beta}^T\mathbf{X}]$ is commonly referred to as the relative risk, and $\gamma$ and $\mat{\beta}$ are unknown parameters commonly known as hazard ratio associated with covariates $W$ and $\mathbf{X}$, respectively.

The unknown parameters are routinely estimated via optimizing the partial likelihood function of the form
\begin{equation}\label{origpartlikefunc}
\prod_{i=1}^n \left[\frac{r_i(\gamma,\mat{\beta})}{\sum_{j\in R_i}r_j(\gamma,\mat{\beta})] }\right]^{\delta_i},
\end{equation}
where $R_i$ is the risk set associated with time $y_i$.

When the $W_i$ covariate is fully observed without censoring, the optimization procedure can be carried out utilizing a variety of readily available statistical packages.

\subsection{Extension of innovation theorem}\label{InnThm}

Although equation \eqref{coxconcept} characterizes the underlying data generation process, which remains true regardless of censoring in either the outcome or the covariate, the presence of censored covariate complicates the estimation of the unknown parameter through equation \eqref{origpartlikefunc}. The challenge is borne out because the relative risk associated with the censored covariate cannot be accurately quantified. This difficulty was previously explored by \cite{Lee03} who quantified the likelihood contribution of the censored covariate through the Innovation Theorem \citep{Anderson93}. Essentially, the relative risk among those patients is replaced by a conditional expectation, $E(r_i(\gamma,\mat{\beta})|W_i>W^c_i, W^c_i=z_i)=\exp[\mat{\beta}^T\mathbf{x}_i]E(\exp[\gamma W_i]|W_i>z_i)$. Instead of deriving an analytical form of the conditional expectation which requires fully a parameterized specification of the distribution of $Z_i$, \cite{Lee03} proposed an empirical weighted average to estimate the conditional expectation. In their proposal the weight is derived from a non-parametric estimate of the survival function of $W_i$, denoted as $\hat{S}(\cdot)$. We adopt the same strategy and propose a unified relative risk formula
\begin{equation}\label{relrisk}
r_i^*(\gamma,\mat{\beta})=\left\{ \begin{array}{lcl}
\exp[\gamma z_i]\exp[\mat{\beta}^T\mathbf{x}_i] & \mbox{if} & \eta_i=1\\
\frac{\sum_{j \in I_{obs}} I(z_j> z_i)\omega_j\cdot \exp[\gamma z_j]}{\sum_{j \in I_{obs}} I(z_j> z_i)\omega_j}\exp[\mat{\beta}^T\mathbf{x}_i]  & \mbox{if} & \eta_i=0\\
\end{array} \right.
\end{equation}
where $I_{obs}=\{j: \eta_j=1\}$, and $\omega_j=\hat{S}(z_j^{-})-\hat{S}(z_j)$ is the weight derived from a nonparametric survival estimator. Several non-parametric survival curve estimators are available, including the Kaplan-Meier, Nelson-Aalen, and Turnbull. Based on our experience, the choice of the non-parametric estimator have no substantial impact on the result.  Therefore, we opt for the Kaplan-Meier estimator due to its popularity and widespread implementation in different statistical packages. Albeit easy to implement, the Kaplan-Meier estimator does not account for fully observed covariates that could improve the survival estimate if a dependency between them is assumed. In that case, we may consider estimating the survival by positing another Cox model to capture the dependency on the observable covariates and constructing the baseline hazard specific to $W$.

We conclude this section by highlighting a potential pitfall of the proposed method. As shown in equation (\ref{relrisk}), the method's success depends on reconstructing the partial likelihood contribution of a censored covariate through a weighted average of qualified patients with observed covariates. While the theoretical foundation is solid, practical implementation requires caution. For example, if a large covariate value is censored at a much smaller value, the weighted average may disproportionately represent smaller observed covariates leading to bias in the likelihood contribution. In addition, if the largest covariate value is censored, the data from that patient cannot be used in the optimization process.

\subsection{Partial likelihood function and its optimization procedure}\label{partlike}
We modify the partial likelihood function in equation (\ref{origpartlikefunc}) by substituting the relative risk with the one defined in equation (\ref{relrisk}). To carry out the optimization, let $y_{(1)}<y_{(1)}<...<y_{(N)}$ be $N$ distinct observed endpoint times, where $j(i)$ is the patient index whose outcome is observed at time $y_{
(i)}$, and $R_i=\{j: y_j>y_{(i)}\}$ is the risk set at time $y_{(i)}$. With these additional notations, the log-partial likelihood can be expressed in the following equivalent form
\begin{equation}\label{propobj}
\mathcal{L}=\sum_{i=1}^N \left[\log r^*_{j(i)}-\log\sum_{k \in R_i}r^*_k \right]
\end{equation}

In the presence of ties in the event time, we adopt Breslow's method to modify the log-partial likelihood. We propose to maximize equation (\ref{propobj}) using the Newton-Raphson method which can be carried out through the following steps:
\begin{enumerate}
  \item Initiation: Set an initial value for $(\tilde{\gamma}~~\tilde{\mat{\beta}}^T)$
  \item Compute the gradient and Hessian of equation (\ref{propobj}) using the formulas in equations \ref{gradpropobj} and \ref{hesspropobj} (in Appendix), respectively.
  \item Update the solution using the formula
  \begin{equation*}
  (\hat{\gamma}~~\hat{\mat{\beta}}^T)=(\tilde{\gamma}~~ \tilde{\mat{\beta}}^T)-\ddot{\mathcal{L}}^{-1}(\tilde{\gamma}, \tilde{\mat{\beta}})\dot{\mathcal{L}}(\tilde{\gamma},\tilde{\mat{\beta}})
  \end{equation*}
  \item Set $(\tilde{\gamma}~~ \tilde{\mat{\beta}}^T)=(\hat{\gamma}~~ \hat{\mat{\beta}}^T)$
  \item Repeat Steps 2 to 4 until convergence.
\end{enumerate}
We refer to the gradient vector $\dot{\mathcal{L}}$ and Hessian matrix $\ddot{\mathcal{L}}$ of the log partial likelihood function in equation (\ref{propobj}), as well those of the relative risk in equation (\ref{relrisk}), detailed in Appendix \ref{app1}.
The convergence criterion in Step 5 is set to be determined by the difference between two consecutive iterations in both the log-partial likelihood and the parameter estimate, as assessed by the infinity norm. We stop the iteration process when these differences are less than $10^{-5}$. At the initiation step, we build another Cox model based on the CC approach and use its estimate as the initial value. From our experience, the convergence criterion is within 10 iterations using this initialization approach.

\section{Simulations}\label{SimSec}

\subsection{Simulation scenarios}
We have designed three simulation scenarios to replicate the data generation process in various real world applications. Each scenario is evaluated over 2,000 repetitions using conventional metrics, including mean absolute bias, empirical standard error, mean squared error (MSE), and empirical 95\% coverage rate. In addition to demonstrating the performance of the proposed estimation procedure, we have included two estimation procedures as references: an ideal situation where all covariates are fully observed and a complete data analysis excluding patients with censored covariates. We introduce the scenarios as follows.

\paragraph*{Scenario 1.}
The first scenario is a direct extension of \cite{Lee03} by adding three additional covariates. The covariate that is subject to censoring is generated from a uniform distribution $U(0,2)$ and its censoring mechanism is another independent $U(0,a)$, where $a$ is chosen such that the censoring rate is controlled at 30\%. The additional fully observable covariates are generated as follows: $X_j, j=1,2$ are $U(0,2)$ with correlation 0.4, and $X_3$ is a multi-nominal of three distinct values $\{1,2,3\}$ with corresponding probabilities $\{0.1,0.3,0.6\}$, respectively. The survival outcome is drawn from an exponential distribution with hazard $\exp[\gamma W+\beta_1 X_1+\beta_2 X_2+\beta_3 I(X_3=2)+\beta_4 I(X_3=3)]$, where the true parameter values are $(\log2,~-\log2,~\log2,~1,~0)$. The censoring time follows another uniform $U(0,b)$ and we control $b$ such that the censoring rate of the outcome is 40\%.

\paragraph*{Scenario 2.}
This scenario is designed to mimic the real clinical applications similar to that in our motivating example, where one of the goals is to investigate if a secondary time-to-event endpoint, PFS, is predictive to the true outcome OS. We intentionally included PFS, a composite endpoint containing death, this allows us to examine how composite time to event endpoints behave in predictive modeling for OS.  To achieve this, we devise this second scenario so that the treatment effect can only impact OS indirectly through PFS. We accomplish this feature by first generating 300 treatment assignments $T$ from a Bernoulli with probability of 0.5, and let PFS be exponentially distributed with expectations of 1.0 and 0.6 for the active $(T=1)$ and standard care $(T=0)$ groups, respectively. OS is generated from an exponential distribution with hazard $\exp[\gamma PFS+\beta I(T=1)]$ and the true parameters are $\gamma=-1/3$ and $\beta=0$. To ensure that the design closely resembles the real application and avoid conflicting censoring status between OS and PFS, we create the censoring variable from a uniform distribution and apply it to both the OS and PFS endpoints. The uniform distribution parameter is chosen such that the censoring rate of the OS and PFS are at 30\% and 20\%, respectively. Lastly, considering death is a common component in both PFS and OS, we randomly select 10\%, 15\% and 20\% of the PFS events and treat them as death and over-write the OS and its censoring status accordingly.

\paragraph*{Scenario 3.}
This scenario is an extension of the simulation design introduced in \cite{Atem19}. We keep all the design elements but adjust the censoring variable. While the censoring variable is generated from Weibull as is in \cite{Atem19}, we fine-tune the Weibull parameters to maintain the censoring rate of the covariate across three different sets of scale and shape parameters. We control the censoring rates of 25\% for both the covariate and outcome. We fix the sample size at 500.

\subsection{Simulation results}
Table \ref{res1} displays the performance of estimating $\gamma$ in the first scenario. When the sample size is 200,
all four methods slightly overestimate the true parameter value, with the CC analysis showing the highest over-estimation, although the magnitude remains small (within the hundredths decimal place). As the sample size increases to 500, the bias decreases further. Given this negligible bias, the variance becomes the main contributor to the MSE, making it the key metric for comparing the performance of the different estimation procedures. Both our proposed method and Atem's approach outperform the CC analysis, as evidenced by the lower MSE. However, a closer examination is necessary to fully understand the relative merits of these methods. As noted by \cite{Atem19}, substituting censored covariate values with imputed ones can lead to an underestimation of the true variability of the parameter estimate. We observed a similar phenomenon in our simulations where the average standard error across 2,000 repetitions was slightly smaller than the empirical standard error. Consequently, the MSE largely driven by the variance component, confirms the superior performance of our proposed method relative to the CC analysis. Furthermore, the empirical coverage rates are close to the nominal 95\% level, confirming that the standard errors derived from the inverse Hessian matrix in the Newton–Raphson optimization provide reliable uncertainty estimates.

\begin{figure}[t]
\centerline{\includegraphics[width=1\textwidth]{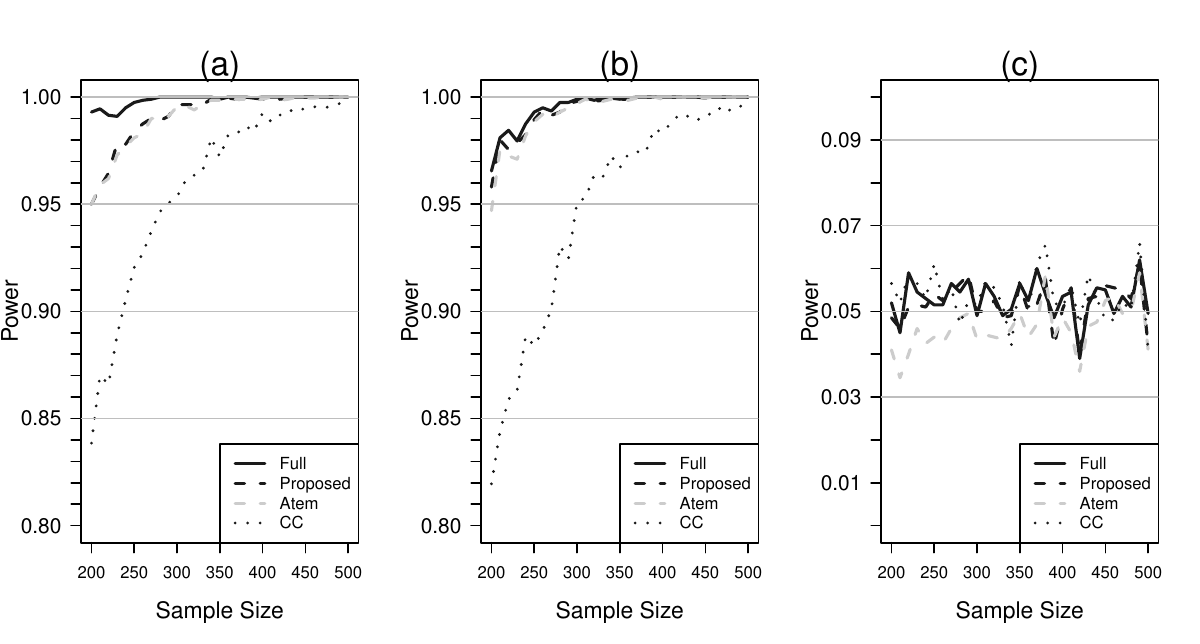}}
\caption{Power associated with testing (a) $\gamma=0$; (b) $\beta_1=0$; and (c) $\beta_4=0$. \label{ex1power}}
\end{figure}

In addition to comparing performance based on bias and variance, we highlight the advantage of the proposed approach from another perspective by examining the empirical power of testing each model parameter at various sample sizes ranging from 200 to 500. The empirical power for testing the nullity of $\gamma, \beta_1$ and $\beta_4$ are shown in panels (a)-(c) of Figure \ref{ex1power}, respectively. While the primary focus of this article is to propose a novel estimation procedure rather than hypothesis testing, these power comparisons provide additional insight into statistical inference and allows us to consider how these methods apply to our motivating example. Each panel illustrates the power comparisons under different contexts. Panel (a) essentially presents the results shown in Table \ref{res1} from a different perspective, without introducing major new information. Although not shown, the performance metrics for estimating $\beta_1$ are not meaningfully different between the Full data and our proposed method (as well \cite{Atem19}). Despite the potential inflation in the covariance of $\hat{\gamma}$ and $\hat{\beta}_1$, there is no substantial efficiency loss in estimating the parameters associated with fully observed covariates. Lastly, the power curves corresponding to $\beta_4$, whose true value is zero, remain flat and fluctuate around the nominal 5\% level across all methods and sample sizes, indicating that the proposed estimation procedure maintains appropriate type I error control. 

Unlike the first simulation scenario, the main objective of the second scenario is not to assess the estimation accuracy but rather to evaluate whether the secondary endpoint is predictive of  OS. By design, OS is a mixture of exponential distributions that do not satisfy the proportional hazards assumption. Consequently, we shift the focus of the simulation and present the results in two parts: qualitative and quantitative evaluation. The qualitative evaluation is based on the empirical frequency with which the parameter estimate of $\gamma$ is statistically significant at the 5\%  level.  In addition, we examine a more stringent condition proposed in \cite{Prentice89}, which requires simultaneously that the intermediate endpoint (PFS) to be predictive of OS while the treatment effect is non-significant in the joint model (third criterion).

\begin{table}[h]
\caption{Simulation Result - Scenario 1.}%
{\begin{tabular}{llcccccc}
\toprule
{Sample Size} & {Method}  & {Average} & {Abs Bias}  & {Emp SE}  & {MSE} & {Emp Coverage} \\
\midrule
200   & Full      &  0.712	& 0.018	& 0.176	& 0.031	& 0.941	 \\
      & Proposed  &  0.707	& 0.014	& 0.204	& 0.042	& 0.944	 \\
      & Atem      &  0.697	& 0.004	& 0.196	& 0.038	& 0.955	 \\
      & CC        &  0.725	& 0.032	& 0.255	& 0.066	& 0.948	 \\
500   & Full      &  0.698	& 0.005	& 0.107	& 0.012	& 0.956	 \\
      & Proposed  &  0.680	& 0.014	& 0.122	& 0.015	& 0.956	 \\
      & Atem      &  0.684	& 0.009	& 0.120	& 0.015	& 0.956	 \\
      & CC        &  0.701	& 0.008	& 0.153	& 0.023	& 0.953	 \\
\bottomrule
	\end{tabular}}\label{res1}
\end{table}

\begin{figure}[h]
\centering
	\includegraphics[width=0.9\textwidth]{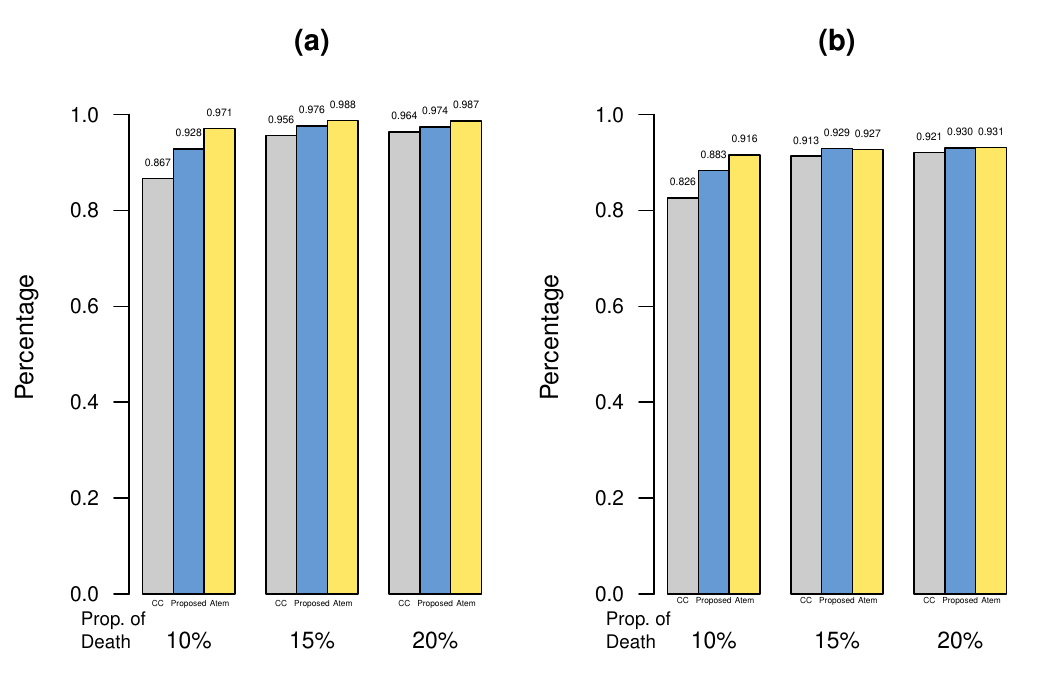}
	\caption{Surrogacy of PFS: (a) PFS is predictive of OS controlling for treatment effect; (b) simultaneous predictive PFS effect and null treatment effect.}
	\label{surr}
\end{figure}

The percentage of satisfying these two conditions are presented in Figure \ref{surr}, panels (a) and (b), respectively. The bar chart in panel (a) resembles the power curve in Figure \ref{ex1power} that shows a superior conditional testing power achieved by the proposed and Atem methods. Moreover, it should be noted that predictability increases as the proportion of deaths increases. While this is intuitively reasonable, this finding reveals a cautionary point when applying and interpreting these methods in real applications. Secondary endpoints, such as PFS, typically include death as a component, which can inevitably inflate their predictability for OS when death precedes more frequently than disease progression. 

\begin{figure}[h]
\centering
	\includegraphics[width=1\textwidth]{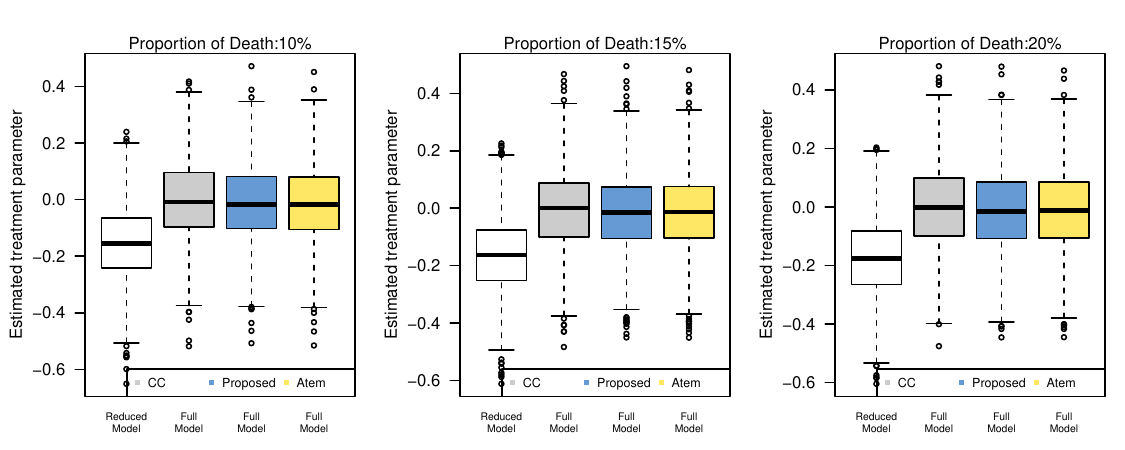}
	\caption{Estimated treatment effect from a full model (with PFS and treatment as covaiate) and a reduced model (treatment only).}
	\label{TrtEstiBox}
\end{figure}

In contrast, the quantitative evaluation examines how the magnitude of the treatment effect, as assessed by the Cox model parameter, changes in the absence versus presence of PFS.  To this purpose, an additional simple Cox model is fitted with treatment assignment as the sole covariate. The parameter estimate associated with treatment effect from these two models are presented in Figure \ref{TrtEstiBox}. While a clear treatment effect on OS is evident in the reduced model (without PFS), the boxplots show that this treatment effect is substantially attenuated after including PFS into the model. Specifically, using the proposed and Atem estimators, the magnitude of treatment effect in the full model is only 7\% to 9\% of that in the reduced model.



\begin{figure}[h]
\centering
	\includegraphics[scale=0.7]{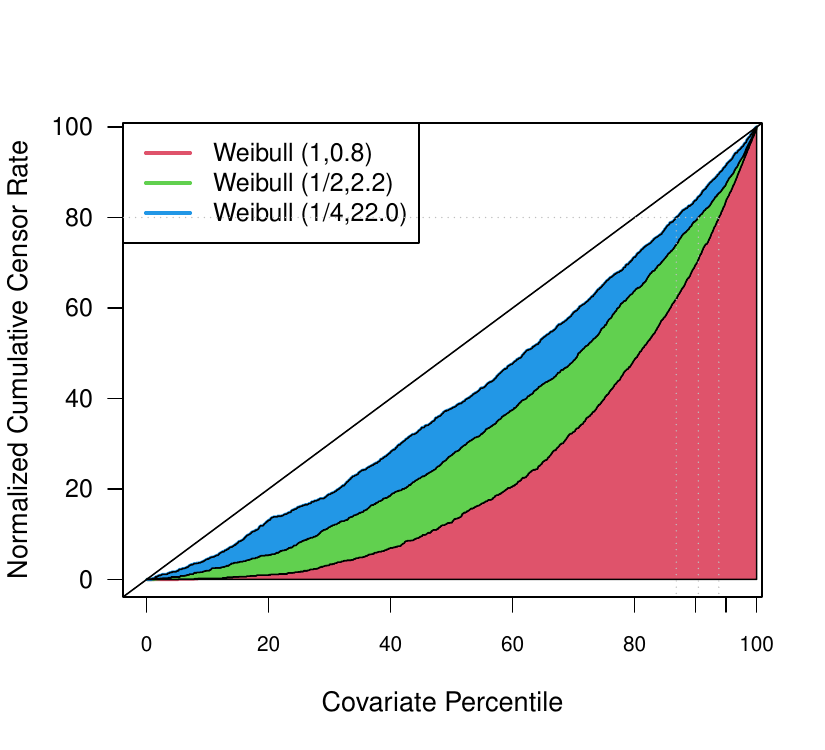}
	\caption{Cumulative covariate censoring rate against covariate percentile by Weibull parameters.}
	\label{Lorenz}
\end{figure}

Before presenting the performance of the third scenario, it is instructive to first highlight how the three censoring mechanisms differ, particularly with respect to the origin of censored covariate. To illustrate this concept, we extend the use of the Lorenz curve, commonly used in economics to visualize equality. In this context, equality suggests each covariate is equally like to be censored independent of its true value. In Figure \ref{Lorenz}, the x-axis is the empirical percentile of the true covariate value; while the y-axis is the cumulative censoring rate normalized to 100\%. If censoring is equally likely across the range, the curve follows a 45-degree straight line. In contrast, when censoring are narrowly clustered toward the large value, the line will be pushed toward the bottom right corner. Among these three settings, the most unequal case is the Weibull(1,0.8), where the smallest 56\% contribute to 1/5 of the censoring, while the top 6\% also contribute 1/5 of the censoring.

We present the estimation performance of the third scenario in Table \ref{res3}. Starting from the most unequal censoring distribution Weibull (1,0.8), the advantages of the proposed method observed in the previous two scenarios are less pronounced. Although the proposed method still outperforms CC data analysis in terms of MSE, it tends to under-estimate the true effect and suffers from under-coverage. As indicated at the end of Section \ref{InnThm}, it is critical to have adequate observed covariates that are both smaller and larger than the true covariate value that is being censored, to ensure the weighted average accurately reconstructs the true likelihood contribution. In this particular setting, given that large covariates are more likely to be censored, it is probable that smaller covariates are over-represented in the weighted average and insufficient larger observed covariate to offset their influence.

Shifting toward a lesser extreme censoring mechanism, the advantages of the proposed method become evident again across all metrics: bias and empirical SE decrease, and coverage rate increases to the nominal level. It is noteworthy that the censoring mechanism also affects the performance of the CC data analysis just as much as the proposed method. Despite the CC data analysis being unbiased across all three censoring distributions, the efficiency loss relative to the Full data analysis decreases considerably from 1.80 times SE under the most unequal case to 1.25 under the least unequal censoring.

\begin{table}[h]
\caption{Simulation Result - Scenario 3.}
{\begin{tabular}{llcccccc}
\toprule
{Censoring Distribution} & {Method}  & {Average} & {Abs Bias}  & {Emp SE}  & {MSE} & {Emp Coverage} \\
\midrule
Weibull(1,0.8)     & Full      & -1.004 & 0.004 & 0.168 & 0.028 & 0.958\\
                   & Proposed  & -1.102 & 0.102 & 0.238 & 0.067 & 0.925\\
                   & Atem      & -1.030 & 0.030 & 0.203 & 0.042 & 0.949\\
                   & CC        & -1.010 & 0.010 & 0.303 & 0.092 & 0.940\\
Weibull(1/2,2.2)   & Full      & -1.004 & 0.004 & 0.168 & 0.028 & 0.958\\
                   & Proposed  & -1.016 & 0.016 & 0.203 & 0.042 & 0.952\\
                   & Atem      & -0.992 & 0.008 & 0.193 & 0.037 & 0.954\\
                   & CC        & -1.005 & 0.005 & 0.234 & 0.055 & 0.946\\
Weibull(1.4,22.0)  & Full      & -1.004 & 0.004 & 0.168 & 0.028 & 0.958\\
                   & Proposed  & -1.005 & 0.005 & 0.199 & 0.040 & 0.948\\
                   & Atem      & -0.989 & 0.011 & 0.192 & 0.037 & 0.945\\
                   & CC        & -1.005 & 0.005 & 0.212 & 0.045 & 0.950\\
\bottomrule
\end{tabular}}\label{res3}
\end{table}

\section{Real data application}\label{realdata}
We applied the proposed estimation procedure to two oncology studies in patients with prostate cancer \citep{chaarted} and breast cancer \citep{rotterdam} aiming to examine the predictability of intermediate time to event endpoints as illustrated in the second simulation scenario.

The prostate cancer trial was previously published in \cite{chaarted}. Briefly, 790 eligible patients were randomized in a 1:1 allocation ratio to either androgen-deprivation (ADT) therapy alone or ADT plus docetaxel. The trial met the primary endpoint of OS demonstrating that patients treated with the combination of ADT and docetaxel had longer survival duration. In this real data analysis, we sought to explore whether both PFS and TTP predict OS. In the CHAARTED trial, ~ 4\% (29/790 patients) had died without progression.

The breast cancer dataset includes nearly 3,000 patients was part of the Rotterdam tumor bank \citep{rotterdam}.
The objective of this breast cancer study was to identify potential predictors based on the urokinase-type plasminogen activator. Our interest is to explore whether relapse-free survival (RFS) and time-to-relapse (TTR) are predictive of OS.

We adapt a hybrid analytical strategy toward the real data analysis, where we first randomly shuffled the treatment assignment to eliminate its effect on OS. Given the absence of treatment assignment in the breast cancer example, we generated a synthetic binary treatment variable from a Bernoulli random variable with probability of 0.5. Additionally, instead of utilizing the full trial participants all at once, we repeatedly sampled 2000 times subsets of 100, 200 and 500 patients, following the same randomization ratio. For each subset, we fit a Cox model analogous to that used in the second simulation setting and then employed three estimation methods, our proposed method, Atem's method and CC analysis, to test the statistical significance of intermediate endpoint while adjusting for the treatment effect. Because treatment assignment has been shuffled (or artificially created), the testing results reflect the true predictive value of the intermediate time to event endpoint, unconfounded by treatment.

\begin{figure}[t]
\includegraphics[width=1\textwidth]{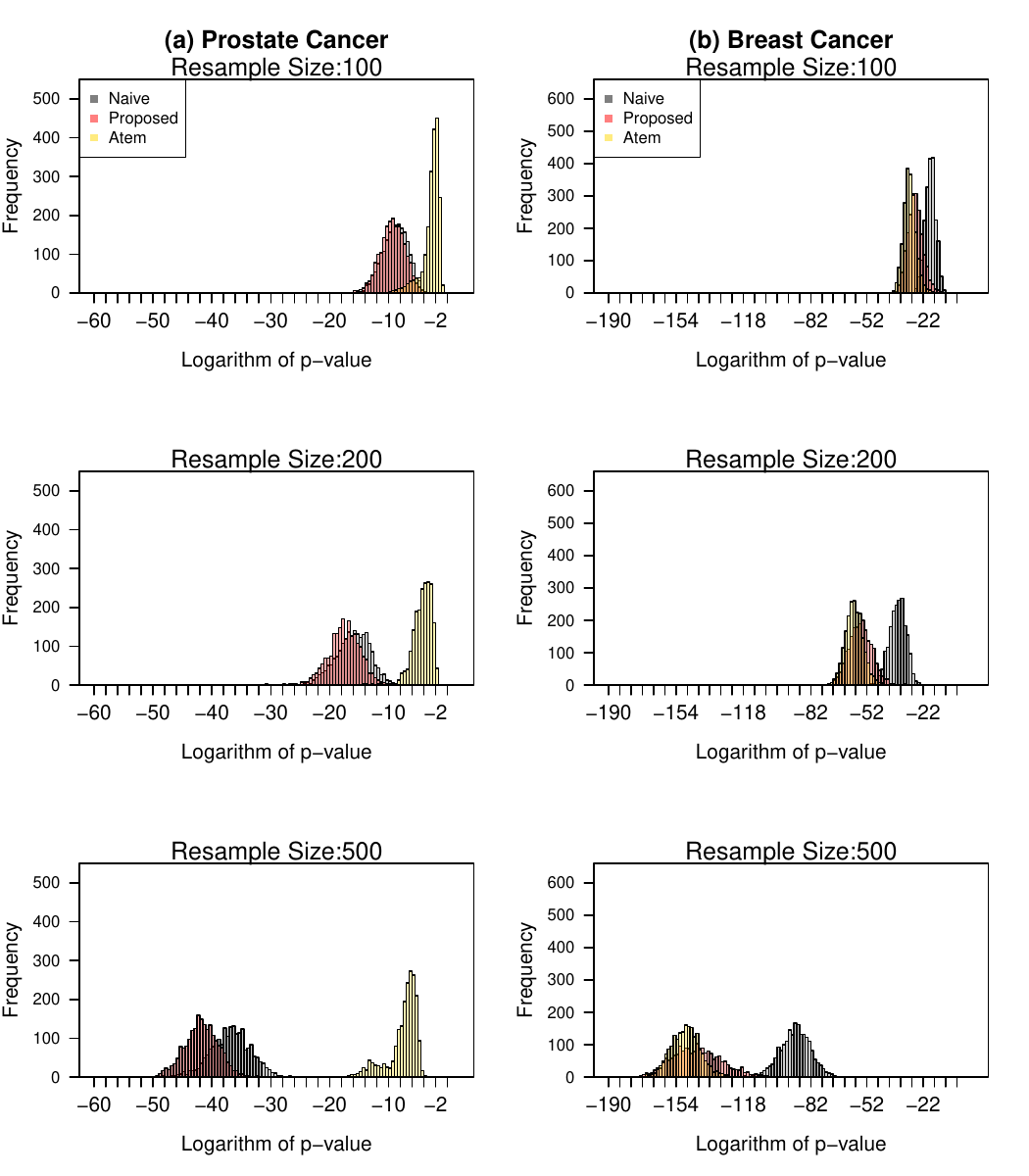}
\caption{(a) Surrogacy of TTP of Prostate Cancer Example; (b) Surrogacy of TTR of Breast Cancer Example.}\label{pvalhist}
\end{figure}

\begin{figure}[t]
\includegraphics[width=1\textwidth]{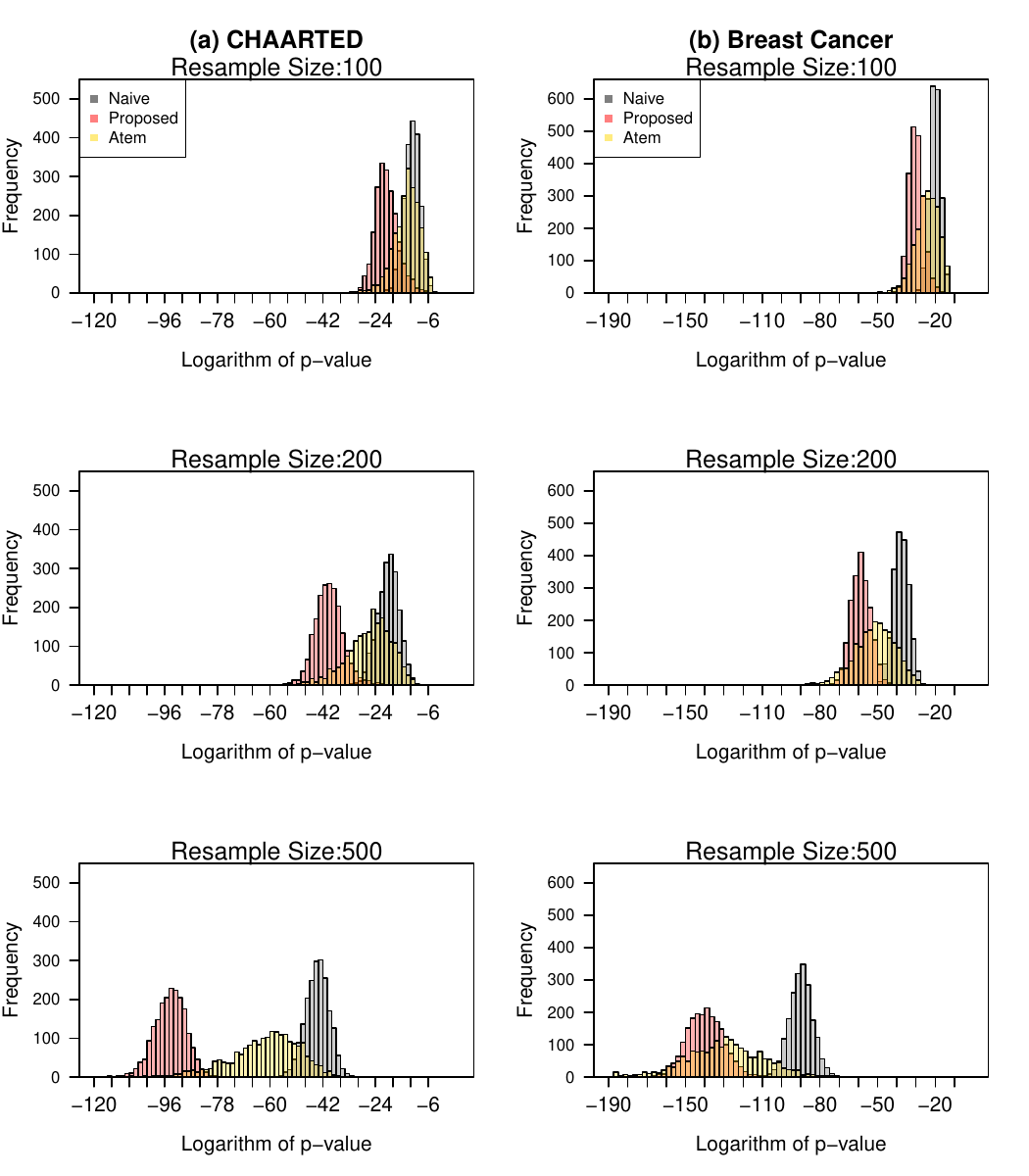}
\caption{(a) Surrogacy of PFS of Prostate Cancer Example; (b) Surrogacy of RFS of Breast Cancer Example.\label{pvalhist2}}
\end{figure}

We show the distribution of 2,000 p-values in logarithm scale in Figure \ref{pvalhist} with the prostate and breast cancer studies on panels (a) and (b), respectively. 
For reference,  Figure \ref{pvalhist2} presents analogous results for PFS and RFS.
Both (TTP, PFS) and (TTR, RFS) are significantly predictive of OS, indicating that these surrogate endpoints behaved similarly.

Consistent with previous findings from meta-analytic results in \cite{metaSurr}, which demonstrated that the PFS is a valid surrogate for OS as quantified by Kendall’s $\tau$ correlation, the histogram in panel (a) indicates that TTP is highly significant even in a subset much smaller than the full participants. As the resample size increases, clear separation emerged between the three histograms, with the p-values from the proposed method being the largest, followed by CC method and Atem's method. In contrast, results from the breast cancer study panel (b) showed somewhat different pattern, with the CC method produced the smallest p-values, while the other methods yield similar small p-values. Moreover, the variability of the p-values is smaller in Atem's approach.
Although the distribution of p-values also suggests strong predictive relationship, it remains uncertain whether time-to-relapse serves as a surrogate for OS in this cohort.

\section{Conclusion}\label{conclusion}
In this article, we introduce a novel estimation method to address the complexity of estimating the hazard ratio in the proportional hazards model when the covariate is subjected to right censoring. This challenge arises because traditional methods fail to account for the incomplete covariate due to censoring, leading potentially to biased estimates. Our approach leverages the Innovation Theorem\citep{Anderson93}, which enables us to quantify the likelihood contribution of patients with censored covariates by utilizing weighted averages of patients with fully observed covariates. By retaining all available information, this approach enhances the robustness and efficiency of the estimation procedure.

Through simulation studies and real data analyses based on two oncology trials \citep{chaarted,rotterdam}, we show that our approach substantially outperforms the naive CC analysis, which discards individuals with censored covariates. Furthermore, our method performs at least as well as one of the existing methods, supporting its potential as new standard for handling censored covariates in practice. 

In addition to proposing this new estimation approach, we also highlight the unique challenges inherent in analyzing censored covariates (this also applies to intermediate time to event endpoints). Unlike censoring in the dependent variable (outcome), the complexity dealing with censored covariates cannot be adequately captured by a single metric, such as the censoring rate. For instance, in our third simulation scenario, although the censoring rates were the same across three different censoring mechanisms, the performance of our approach deteriorates as the inequality in censoring increased. This finding underscores the fact that while the simulation enables us to portray inequality in censoring, it is important to note that detecting such phenomenon is virtually impossible in practice. Therefore, we recommend that researchers create a histogram of the observed covariate as a diagnostic tool, with nonlinear drops in the histogram bars, and/or unexpectedly low upper limits should prompt careful scrutiny of the results. Moreover, as illustrated in Figure \ref{ex1power}, the superior estimation precision of our approach translates into a reduced sample size requirement compared to the CC approach, particularly as the covariate censoring rate increases.  By efficiently utilizing all available information, our approach preserves statistical power and mitigates the data loss inherent to CC analyses.

While our primary focus is on developing this estimation method, it can also be readily applied to the identification of surrogate endpoints in clinical research. Importantly, our objective is not to claim causality but to quantify the predictive surrogacy of intermediate time-to-event outcomes for OS. This framework enables a formal evaluation of the predictive utility of intermediate endpoints in accordance with the Prentice criteria \citep{Prentice89}, which require that a valid surrogate fully captures the effect of treatment on the true endpoint (criterion 3). By modeling the relationship between the intermediate endpoint and OS while adjusting for covariates and appropriately handling censoring, our method provides a rigorous and quantitative assessment of whether an intermediate time-to-event outcome can serve as a reliable surrogate for survival. Crucially, findings from such analyses should not be interpreted as evidence that a treatment prolonging an intermediate endpoint (such as TTP) necessarily extends OS, but rather as an assessment of whether the intermediate outcome provides a robust measure of treatment effect.


In conclusion, our proposed method improves estimation accuracy and efficiency in the presence of censored covariates, while providing a principled approach for assessing intermediate time-to-event endpoints as potential surrogates of the true endpoint. By explicitly addressing the unique challenges of censored covariates, our approach enhances predictive modeling, strengthens surrogate identification, and supports robust statistical inference in clinical research.

\backmatter







\section*{Declarations}

\bmhead{Conflict of interest}
The authors declare no potential conflict of interests.

\bmhead{Funding}
This research was supported in part from the National Institutes of Health grants R01 CA256157, R01 CA249279, R21 CA263950, R01 LM013352, the United States Army Medical Research Materiel Command awards  HT9425-23-1-0393 and HT9425-25-1-0623, the Food and Drug Administration (FDA) Award U01 FD007857 of the U.S. Department of Health and Human Services (HHS), and the Prostate Cancer Foundation Challenge Award. The content is solely the responsibility of the authors and do not represent the official views of, nor an endorsement, by FDA/HHS, or the U.S. Government. Taehwa Choi acknowledges partial support from the National Research Foundation of Korea (NRF) grant funded by the Korean government (RS-2024-00340298).

\bmhead{Data availability}
The clinical data from the CHAARTED trial can be accessed through the NCTN NCORP Data Archive. The breast cancer dataset is publicly available at R survival package \citep{Rcran}.

\begin{appendices}

\section{Gradient vector and Hessian matrix of the partial likelihood}\label{app1}
The optimization of the log partial likelihood in equation (\ref{propobj}) requires the gradient vector and Hessian matrix of both the log partial likelihood itself and the relative risk in equation (\ref{relrisk}) with respect to the unknown parameter $(\gamma~~\mat{\beta}^T)$. We begin with those of the log partial likelihood, denoted as $\dot{\mathcal{L}}$ and $\ddot{\mathcal{L}
}$,
\begin{equation}\label{gradpropobj}
\dot{\mathcal{L}}=\sum_{i=1}^N \left[\frac{\triangledown r_{j(i)}^*} {r^*_{j(i)} }-\frac{\sum_{k \in R_i}\triangledown r^*_k}{\sum_{k \in R_i}r^*_k} \right],
\end{equation}
 and
\begin{equation}\label{hesspropobj}
\ddot{\mathcal{L}}=\sum_{i=1}^N \left[\frac{\triangledown^2 r_{j(i)}^*} {r^*_{j(i)} }- \left( \frac{\triangledown r_{j(i)}^*} {r^*_{j(i)} } \right)\left( \frac{\triangledown r_{j(i)}^*} {r^*_{j(i)} } \right)^T\right]-\left[\frac{\sum_{k \in R_i}^2 r^*_k}{\sum_{k \in R_i}r^*_k}-\left( \frac{\sum_{k \in R_i}\triangledown r^*_k}{\sum_{k \in R_i}r^*_k} \right)\left(\frac{\sum_{k \in R_i}\triangledown r^*_k}{\sum_{k \in R_i}r^*_k}\right)^T\right],
\end{equation}
where $\triangledown r^*$ and $\triangledown^2 r^*$ are the gradient vector and Hessian matrix of the relative risk given by

\begin{equation}\label{gradrr}
\triangledown r_i^*=\left\{ \begin{array}{lcl}
\exp[\mat{\beta}^T\mathbf{x}_i]\exp[\gamma z_i]\left(\begin{array}{c}z_i \\ \mathbf{x}_i\end{array}\right) & \mbox{if} & \eta_i=1\\
\exp[\mat{\beta}^T\mathbf{x}_i]\left(\begin{array}{l} \frac{\sum_{j \in I_{obs}} I(z_j> z_i)W_j\cdot \exp[\gamma z_j]z_j}{\sum_{j \in I_{obs}} I(z_j>
z_i)\omega_j}\\ \frac{\sum_{j \in I_{obs}} I(z_j> z_i)\omega_j\cdot \exp[\gamma z_j]}{\sum_{j \in I_{obs}} I(z_j>
z_i)\omega_j}\mathbf{x}_i\end{array}\right)
  & \mbox{if} & \eta_i=0\\
\end{array} \right.
\end{equation}
and

\begin{equation}\label{hessrr}
\triangledown^2 r_i^*=\left\{ \begin{array}{lcl}
\exp[\mat{\beta}^T\mathbf{x}_i]\exp[\gamma z_i]\left(\begin{array}{c}z_i \\ \mathbf{x}_i\end{array}\right)(z_i~~\mathbf{x}^T_i) & \mbox{if} & \eta_i=1\\
\exp[\mat{\beta}^T\mathbf{x}_i]\left(\begin{array}{ll}  \frac{\sum_{j \in I_{obs}} I(z_j> z_i)\omega_j\cdot \exp[\gamma z_j]z_j^2}{\sum_{j \in I_{obs}} I(z_j>
z_i)\omega_j} & \frac{\sum_{j \in I_{obs}} I(z_j> z_i)\omega_j\cdot \exp[\gamma z_j]z_j}{\sum_{j \in I_{obs}} I(z_j>
z_i)\omega_j}\mathbf{x}_i^T \\ \frac{\sum_{j \in I_{obs}} I(z_j> z_i)\omega_j\cdot \exp[\gamma z_j]z_j}{\sum_{j \in I_{obs}} I(z_j>
z_i)\omega_j}\mathbf{x}_i & \frac{\sum_{j \in I_{obs}} I(z_j> z_i)\omega_j\cdot \exp[\gamma z_j]}{\sum_{j \in I_{obs}} I(z_j>
z_i)\omega_j}\mathbf{x}_i\mathbf{x}_i^T \end{array}\right)
 & \mbox{if} & \eta_i=0
\end{array}\right.
\end{equation}




\end{appendices}


\bibliography{sn-bibliography}

@article{metaSurr,
author = {Halabi, Susan and Roy, Akash and Rydzewska, Larysa and Godolphin, Peter and Parmar, Mahesh K. B. and Hussain, Maha H. A. and Tangen, Catherine and Thompson, Ian and Xie, Wanling and Carducci, Michael Anthony and Smith, Matthew Raymond and Morris, Michael J. and Gravis, Gwenaelle and Dearnaley, David P. and Verhagen, Paul and Goto, Takayuki and James, Nicholas D. and Buyse, Marc E. and Tierney, Jayne F. and Sweeney, Christopher},
title = {Assessing intermediate clinical endpoints (ICE) as potential surrogates for overall survival (OS) in men with metastatic hormone-sensitive prostate cancer (mHSPC).},
journal = {Journal of Clinical Oncology},
volume = {40},
number = {16\_suppl},
pages = {5006-5006},
year = {2022},
doi = {10.1200/JCO.2022.40.16\_suppl.5006},
}

@article{Matsubara20,
author = {Matsubara, N and Chi, KN and Özgüroğlu, M and Rodriguez-Antolin, A and Feyerabend, S and Fein, L and Alekseev, BY and Sulur, G and Protheroe, A and Li, S and Mundle, S and De Porre, P and Tran, N and Fizazi, K},
title = {Correlation of Prostate-specific Antigen Kinetics with Overall Survival and Radiological Progression-free Survival in Metastatic Castration-sensitive Prostate Cancer Treated with Abiraterone Acetate plus Prednisone or Placebos Added to Androgen Deprivation Therapy: Post Hoc Analysis of Phase 3 LATITUDE Study},
journal = {Eur Urol},
volume = {77},
number = {4},
pages = {494-500},
year = {2020},
doi = {10.1016/j.eururo.2019.11.021.},
}

@article{Chowdhury23,
author = {Chowdhury, S and Bjartell, A and Agarwal, N and Chung, BH and Given, RW and Pereira de Santana Gomes, AJ and Merseburger, AS and Özgüroğlu, M and Ju\'arez Soto, \'A and Uemura, H and Ye, D and Brookman-May, SD and Londhe, A and Bhaumik, A and Mundle, SD and Larsen, JS and McCarthy, SA and Chi, KN},
title = {Deep, rapid, and durable prostate-specific antigen decline with apalutamide plus androgen deprivation therapy is associated with longer survival and improved clinical outcomes in TITAN patients with metastatic castration-sensitive prostate cancer},
journal = {Ann Oncol},
volume = {34},
number = {5},
pages = {477-485},
year = {2023},
doi = {10.1016/j.annonc.2023.02.009},
}

@article{Cortazar14,
author = {Cortazar, P and Zhang, L and Untch, M and Mehta, K and Costantino, JP and Wolmark, N and Bonnefoi, H and Cameron, D and Gianni, L and Valagussa, P and Swain, SM and Prowell, T and Loibl, S and Wickerham, DL and Bogaerts, J and Baselga, J and Perou, C and Blumenthal, G and Blohmer, J and Mamounas, EP and Bergh, J and Semiglazov, V and Justice, R and Eidtmann, H and Paik, S and Piccart, M and Sridhara, R and Fasching, PA and Slaets, L and Tang, S and Gerber, B and Geyer, CE and Pazdur, R and Ditsch, N and Rastogi, P and Eiermann, W and von Minckwitz, G},
title = {Pathological complete response and long-term clinical benefit in breast cancer: the CTNeoBC pooled analysis},
journal = {Lancet},
volume = {384},
number = {9938},
pages = {164-172},
year = {2014},
doi = {10.1016/S0140-6736(13)62422-8},
}

@article{Halabi09,
author = {Halabi, S and Vogelzang, NJ and Ou, SS and Owzar, K and Archer, L and Small, EJ},
title = {Progression-free survival as a predictor of overall survival in men with castrate-resistant prostate cancer},
journal = {Journal of Clinical Oncology},
volume = {27},
number = {17},
pages = {2766-2771},
year = {2009},
doi = {10.1200/JCO.2008.18.9159},
}

@article{Halabi08,
author = {Halabi, S and Vogelzang, NJ and Kornblith, AB and Ou, SS and Kantoff, PW and Dawson, NA and Small, EJ},
title = {Pain predicts overall survival in men with metastatic castration-refractory prostate cancer},
journal = {Journal of Clinical Oncology},
volume = {26},
number = {15},
pages = {2544-2549},
year = {2008},

}

@article{Lee03,
      author = {Lee, S and Park, SH and Park, J},
       title = {The proportional hazards regression with a censored covariate},
     journal = {Statistics and Probability Letters},
        year = {2003},
      volume = {61},
       pages = {309--319}
}

@article{Prentice89,
      author = "Prentice, RL",
       title = "Surrogate endpoints in clinical trials: definition and operational criteria",
        year = "1989",
     journal = "Statistics in Medicine",
      volume = "8",
      number = "",
       pages = {431--440}
}

@article{chaarted,
author = {Sweeney, Christopher J. and Chen, Yu-Hui and Carducci, Michael and Liu, Glenn and Jarrard, David F. and Eisenberger, Mario and Wong, Yu-Ning and Hahn, Noah and Kohli, Manish and Cooney, Matthew M. and Dreicer, Robert and Vogelzang, Nicholas J. and Picus, Joel and Shevrin, Daniel and Hussain, Maha and Garcia, Jorge A. and DiPaola, Robert S.},
title = {Chemohormonal Therapy in Metastatic Hormone-Sensitive Prostate Cancer},
journal = {New England Journal of Medicine},
volume = {373},
number = {8},
pages = {737-746},
year = {2015} }

@article{rotterdam,
      author = "Foekens, J and Peters, H and Look, M and Portengen, H and Schmitt, M and Kramer, M and Brunner, N and Jänicke, F and Meijer-van Gelder, M and Henzen-Logmans, S and van Putten, W and Klijn, J",
       title = "The urokinase system of plasminogen activation and prognosis in 2780 breast cancer patients",
        year = "2000",
     journal = "Cancer Research",
      volume = "60",
      number = "",
       pages = "636:643"
}

@Manual{Rcran,
    title = {R: A Language and Environment for Statistical Computing},
    author = {{R Core Team}},
    organization = {R Foundation for Statistical Computing},
    address = {Vienna, Austria},
    year = {2022},
    url = {https://www.R-project.org/},
  }

@article{Xie2020,
author = {Xie, W and Regan, MM. and Buyse, M and Halabi, S and Kantoff, PW. and Sartor, O and Soule, H and Berry, D and Clarke, N and Collette, L and D’Amico, A and Lourenco, RDA and Dignam, J and Eisenberger, M and James, N and Fizazi, K and Gillessen, S and Loriot, Y and Mottet, N and Parulekar, W and Sandler, H and Spratt, DE and Sydes, MR and Tombal, B and Williams, S and Sweeney, CJ},
title = {Event-Free Survival, a Prostate-Specific Antigen–Based Composite End Point, Is Not a Surrogate for Overall Survival in Men With Localized Prostate Cancer Treated With Radiation},
journal = {Journal of Clinical Oncology},
volume = {38},
number = {26},
pages = {3032-3041},
year = {2020},
doi = {10.1200/JCO.19.03114},
}

@article{Qian18,
      author = "Qian, J and Chiou, SH and Maye, JE and Atem, F and Johnson, KA and Betensky, RA",
       title = "Threshold regression to accommodate a censored covariate",
        year = "2018",
     journal = "Biometrics",
      volume = "74",
      number = "4",
       pages = "1261--1270"
}

@article{Tsimikas12,
      author = "Tsimikas, J.V. and Bantis, L.E. and Georgiou, S.D.",
       title = "Inference in generalized linear regression models with a censored covariate",
        year = "2012",
     journal = "Computational Statisticsd and Data Analysis",
      volume = "56",
      number = "6",
       pages = "1854--1868"
}

@article{Atem17,
      author = "Atem, FD and Qian, J and Maye, JE and Johnson, KA and Betensky, RA",
       title = "Linear regression with a randomly censored covariate: application to an Alzheimer’s study",
        year = "2017",
     journal = "J Royal Stat Soc Ser C",
      volume = "66",
      number = "2",
       pages = "313--328"
}

@article{Bernhardt14,
      author = "Bernhardt, P W and Wang, J W and Zhang, D.",
       title = "Flexible modeling of survival data with covariates subject to detection limits via multiple imputation",
        year = "2014",
     journal = "Computational Statisticsd and Data Analysis",
      volume = "69",
      number = "",
       pages = "81--91"
}

@article{Matsouaka20,
      author = "Matsouaka, F D and Atem, F D",
       title = "Regression with a right-censored predictor using inverse probability weighting methods",
        year = "2020",
     journal = "Statistics in Medicine",
      volume = "39",
      number = "",
       pages = "3001--4015"
}

@article{Atem19,
      author = "Atem, F D and Matsouaka, R A and Zimmern, V E ",
       title = "Cox regression model with randomly censored covariates",
        year = "2019",
     journal = "Biometrical Journal",
      volume = "61",
      number = "4",
       pages = "1020--1032"
}

@article{DAngelo08,
      author = "D'Angelo, G. and Weissfeld, L. and GenIMS Investigators",
       title = "An index approach for the Cox model with left censored covariates",
        year = "2008",
     journal = "Statistics in medicine",
      volume = "27",
      number = "22",
       pages = "4502--4514"
}

@book{Anderson93,
      author = "Anderson, P K and Borgan, O and Gill, R D and Keiding, N",
       title = "Statistical {M}odels {B}ased on {C}ounting {P}rocess",
   publisher = "Springer-Verlag",
     address = "New York",
        year = "1993",
        note = "ISBN 978-1-4612-4348-9"
}

@book{SurvBook,
      author = "Klein, JP and  Moeschberger, ML ",
       title = "Survival analysis: Techniques for censored and truncated data",
   publisher = "Springer",
     address = "New York",
        year = "2003",
        note = "ISBN 978-1-4419-2985-3"
}

@book{MissingBook,
      author = "Little, R.J.A. and  Rubin, D B",
       title = "Statistical Analysis with Missing Data",
   publisher = "Wiley",
   address = "New Jersey",
        year = "2019",
        note = "ISBN 978-0-470-52679-8"
}

\end{document}